\documentclass[final,5p,times,twocolumn,authoryear]{elsarticle}

\usepackage{amssymb}
\usepackage{amsmath}
\usepackage{lipsum}
\usepackage{booktabs}
\usepackage{multirow}

\journal{New Astronomy}

\begin{document}

\begin{frontmatter}



\title{Evolution of neutron stars in wide eccentric low-mass binary systems}


\author[1,2]{Marina Afonina\corref{cor1}}
\ead{afonina.md19@physics.msu.ru}
\author[1]{Sergei Popov}
\ead{sergepolar@gmail.com}

\affiliation[1]{organization={Sternberg Astronomical Institute, Lomonosov Moscow State University},
            addressline={Universitetskij pr. 13}, 
            city={Moscow},
            postcode={119234},
            country={Russia}}
\affiliation[2]{organization={Department of Physics, Lomonosov Moscow State University},
            addressline={1/2 Leninskie Gory}, 
            city={Moscow},
            postcode={119991},
            country={Russia}}
\cortext[cor1]{Corresponding author}

\begin{abstract}
Precise astrometric measurement with {\it Gaia} satellite resulted in the discovery of tens of wide binary systems consisting of a Sun-like star and an invisible component. The latter can be a white dwarf, a neutron star, or a black hole.   
In this paper, we model magneto-rotational evolution of neutron stars in wide low-mass binaries accounting for the orbital eccentricity. We aim to calculate when neutron stars in such systems can start to accrete matter from the stellar wind of the companion.
We show that the transition from the ejector to the propeller stage occurs earlier in more eccentric systems, thus increasing the time that neutron stars can spend accreting matter.
Our calculations show that in the case of efficient spin-down at the propeller stage, a neutron star in an eccentric orbit with $e\gtrsim0.6$ and a standard magnetic field $B=10^{12}$~G can start accreting within a few Gyr. For neutron stars with $B=10^{13}$~G the onset of accretion occurs earlier regardless of the orbital eccentricity.
Otherwise, with a lower spin-down rate, such a neutron star will remain at the propeller stage for most of its life.

\end{abstract}




\end{frontmatter}




\section{Introduction}
\label{introduction}

 White dwarfs (WDs), neutron stars (NSs), and stellar mass black holes (BHs) are end-points of stellar evolution. There are about a hundred million BHs, up to a billion NSs, and several billion WDs in the Galaxy \citep{2010A&A...523A..33S, 2020ApJ...889...31L}. There are many methods to discover these objects. Some of them are based on the direct observation of their emission (e.g., in the case of young hot WDs and NSs). Many compact objects are identified due to accretion in binary systems. Still, the majority of stellar remnants avoid detection as they are dim by themselves and do not accrete. 
 
Being massive, compact objects in specific cases might become observable due to their gravity. There are dozens of examples when isolated compact objects are spotted due to microlensing (see e.g., \citet{2021arXiv210713697M, 2024MNRAS.531.2433S} and references therein). This number is expected to grow when new {\it Gaia} data releases are published. The future {\it Roman} space telescope is expected to be very effective in discovering compact objects as gravitational lenses \citep{2024arXiv240706484G}. 

 However, there are not so many hopes of observing the lensing compact object directly. Still, there is also an old method to identify an invisible massive body via its gravitational influence if it is a member of a binary system. This potentially allows for follow-up observations of a compact object. Already in the XIX century,  
 \cite{1844MNRAS...6R.136B} 
 has discovered massive invisible companions for Sirius and Procyon observing the motion of the visible stars. Later, the massive invisible bodies were found to be WDs.
 The same astrometric approach is also applicable to systems with NSs and BHs. However, as these compact objects are much less abundant than WDs and many systems are destroyed after an NS or a BH formation due to mass loss and kick, the probability of finding a binary with such a compact object is rather low. That is why only recently the first examples of astrometric discoveries of `dormant' NSs and BHs in binary systems appeared \citep{2024NewAR..9801694E}. 

 It is very difficult to identify an invisible component with astrometric observations in the case of close binaries. 
 Then, another approach can help. 
In the mid-1960s, \cite{1966SvA....10..251G} proposed that a BH can be identified as an invisible component in a spectroscopic binary. 
However, only recently has it become possible to identify NSs and BHs as invisible components in such systems (e.g., \cite{2024ApJ...964..101Z} and references therein).

Astrometric and spectroscopic methods allowed the researchers to discover several binaries with 
non-detected compact objects. In this paper, we focus on wide binaries consisting of an NS and a Sun-like star. Examples of such systems can be found in the papers \cite{2022arXiv220700680A, 2023ApJ...944L...4L, 2024OJAp....7E..58E, 2024A&A...686A.299S}. 
We are interested in the possibility of such an NS to become an accreting source since, potentially in this case, electromagnetic emission from the compact object can be detected. 

Accretion can start even from a weak stellar wind of the Sun-like component if the NS can spin down sufficiently and the magnetic field does not prevent penetration of the gravitationally captured matter down to the surface. We perform calculations of the NS spin evolution on an eccentric orbit for constant and decaying magnetic fields. Several models of the propeller stage are applied, as the assumptions about the spin-down rate at this stage drastically influence the possibility of reaching the stage of accretion. 

{ This note is a continuation of our studies of evolution of NSs in wide binary systems. In the first paper
\citep{2024Univ...10..205A}, we applied various propeller models to analyse magneto-rotational evolution of NSs in symbiotic X-ray binaries. In these systems an NS is accreting from the wind of a red giant companion \citep{2019MNRAS.485..851Y,2024ApJ...969...35H}. 
It is interesting to note that some of wide low-mass systems with NSs, similar to those studied below, in future can become symbiotic X-ray binaries \citep{2024PASP..136g4202N}.
In the second paper
\citep{2024arXiv240900714A} we modeled the NS evolution in a wide binary with a solar-like companion in a circular orbit. Here we extend our approach to eccentric systems.
}

The paper is structured as follows. In the next section, we describe our model of magneto-rotational evolution and the assumptions about the stellar wind and binary properties.  
Then in Sec.~3, we present the results of our calculations. Sec.~4 contains the discussion of our assumptions and results. In the final section, we summarize our conclusions.






\section{Model}

In this section, we present the model that we apply to model the evolution of NSs in wide low-mass binaries.
We are mainly interested in the time of accretion onset when the NS can be seen e.g., as an X-ray source. Prior to accretion, the NS interacts with the stellar wind of the second component, changing its spin period and observational appearance. To account for these processes, we first build a model of the spin evolution of the NS, then consider the parameters of the stellar wind, and finally present the full algorithm for calculating the spin evolution on the eccentric orbits.

\subsection{Main parameters}

To describe the spin evolution of an NS we follow the general approach presented e.g., by \cite{1992ans..book.....L, 2024Galax..12....7A}. In this consideration, an NS is characterized by the mass $M$, the rotational frequency $\omega$ (or spin period $P$), and the dipole magnetic field at the equator $B$ (or magnetic moment $\mu=BR_\text{NS}^3$, where $R_\text{NS}$ is the NS radius that we accept to be equal to 10~km). 
The external material is usually described by the velocity $v_\infty$ and the matter density $\rho$. Then, we can define the rate of matter capturing $\dot{M}$. In our model, it is derived from the solution of the Bondi-Hoyle-Lyttleton (BHL) problem (see \citet{2004NewAR..48..843E} for a review).
In the binary, several components of the velocity contribute to $v_\infty$. They are the stellar wind speed $v_\text{w}$, the orbital velocity, and the speed of sound $c_s$. The first two can be combined into the matter flow velocity relative
 to the NS, $v$. Then, the characteristic velocity $v_\infty=\sqrt{v^2+c_\text{s}^2}$. We assume that $c_\text{s}$ is insignificant, and therefore adopt 
\begin{equation}
    \dot{M} = \alpha \pi (GM)^2 \rho v_\infty^{-3}
\end{equation}
with the dimensionless coefficient $\alpha=4$. Now that we have defined all the necessary parameters, we can begin to consider the evolution of the NS.

\subsection{Magneto-rotational evolution of a neutron star}

The NS is usually born as an ejecting pulsar and then spins down losing rotational energy until it starts interacting with the outer material, which later leads to the onset of accretion. Each type of interaction with matter corresponds to a distinct evolutionary stage. There are three main stages: ejector (E), propeller (P), and accretor (A).
In this subsection, we first consider the characteristic radii necessary to determine the evolutionary stage, then define the corresponding spin-down rates, and finally, discuss the magnetic field evolution.

Since the NS is surrounded by the material, there is a characteristic radius, where the matter is influenced by the present mass. It is the gravitational capture (or Bondi) radius, formally the distance where the parameter $v_\infty$ equal to the escape velocity
\begin{equation}
    R_\text{G} = \frac{2GM}{v_\infty^2}.
\end{equation}
We assume that the magnetic field of the NS rotates with the same angular frequency as that of the NS. 
Then, there is a maximum radius at which closed magnetic field lines can exist. This is the light cylinder radius 
\begin{equation}
    R_\text{l} = \frac{c}{\omega}.
\end{equation}

Another important radius is the position of the centrifugal barrier for the material in the magnetosphere with solid body rotation. It is proportional to the corotation radius $R_\text{co}$, where an orbital frequency is equal to $\omega$.
We follow \cite{2023MNRAS.520.4315L} in defining this parameter:
\begin{equation}
    R_\text{cb} = 0.87 R_\text{co} = 0.87 \left(\frac{GM}{\omega^2}\right)^{1/3}.
\end{equation}

The NS interacts with the external medium via either the magnetosphere or the pulsar wind. Early in the evolution, it is assumed that the wind produced by the NS prevents the outer material from entering the magnetosphere beyond $R_\text{G}$. In this case, the pressure balance occurs at the Shvartsman radius
\begin{equation}
     R_\text{Sh}= \left( \frac{\xi \mu^2 \omega^4}{4 \pi  c^4 \rho v_\infty^2}\right)^{1/2},
\end{equation}
where $\xi=2$ \citep{2024PASA...41...14A}.
If there is a probability that the wind of relativistic particles will be produced even when the material is gravitationally captured, then the expression is different. In our consideration, it can form an envelope around the NS with a shallow density profile, as considered by \cite{1981MNRAS.196..209D}. Thus the Shvartsman radius for the gravitationally captured matter is
\begin{equation}
    R_\text{Sh}^\text{env} = R_\text{G}\left(\frac{\xi\mu^2\omega^4}{c^4\dot{M}v_\infty}\right)^2.
\end{equation}

Now we consider how the magnetosphere interacts with material. The Alfven radius is the distance where the ram pressure of free-falling matter is balanced by the magnetic field pressure
\begin{equation}
    R_\text{A} = \left(\frac{\mu^2}{{2 \dot{M} \sqrt{2GM}}}\right)^{2/7}.
\end{equation}
If the NS is surrounded by a shell of heated material whose density profile is flatter than that of the free-falling matter, the magnetospheric radius will be larger \citep{1981MNRAS.196..209D}
\begin{equation}
    R_\text{m} = R_\text{A}^{7/9}R_\text{G}^{2/9}.
\end{equation}
Beyond $R_\text{G}$, the magnetosphere interacts with the matter that is not gravitationally captured. The ram pressure is independent of the distance from the NS, so $R_\text{m}$ is different
\begin{equation}
    R_\text{m} = \left(\frac{\mu^2 R_\text{G}^{2}}{2 \dot{M} v_\infty}\right)^{1/6}.
\end{equation}
The magnetospheric radius cannot physically exceed the light cylinder radius $R_\text{l}$, so we limit $R_\text{m}$ by $R_\text{l}$ in our calculations. 

During the evolution, the critical radii change, which causes the NS stages to switch. At each stage, the evolution of the spin period can be described by the Euler equation
\begin{equation}
\label{euler}
    I \frac{d\omega}{dt} = -K,
\end{equation}
where the spin-down torque $K$ is defined by the current evolutionary stage.

The NS is typically born at the ejector stage. Here, it loses its rotational energy via generating the wind of relativistic particles with the rate $L\sim\mu^2\omega^4/c^3$, so the spin-down torque is
\begin{equation}
\label{K_E}
    K_\text{E} = \xi \frac{\mu^2}{R_\mathrm{l}^3},
\end{equation}
where the dimensional coefficient $\xi=2$. Usually, this regime is kept while $R_\text{Sh} > R_\text{G}$. After $R_\text{Sh}$ becomes smaller, the spin-down with $K_\text{E}$ ends \citep{1992ans..book.....L}. However, it is true only if the matter can fall freely. We argue that due to the shallow pressure profile in the envelope, the pulsar wind can be generated and periodically sweep matter out even if it takes place within $R_\text{G}$. It is the transient ejector stage \citep{2024arXiv240900714A}, where it is assumed that the spin-down rate is similar to the ejector one. The propeller stage begins when the wind cannot balance any more the external pressure neither at the outer boundary of the envelope $R_\text{G}$ (it is represented by the condition $R_\text{Sh} < R_\text{G}$), nor at $R_\text{l}$, where the wind is generated ($R_\text{Sh}^\text{env} < R_\text{l}$).

At the propeller stage, the centrifugal barrier prevents the material from falling onto the surface of the NS (it corresponds to the condition $R_\text{m}>R_\text{cb}$). The NS loses its rotational energy through the inner boundary of the rarefied envelope, where the magnetosphere interacts with the material. The spin-down rate depends on the physical model of the interaction. Table~\ref{tab_prop} lists several models of propeller spin-down in the order starting from the most effective with the highest rate of rotational losses (model A) to the least effective (model D).

\begin{table}[t]
    \centering
    \renewcommand{\arraystretch}{1.4}
    \begin{tabular}{cl}
    \hline
         Model & Braking torque $K$ \\ \hline
         A & $\Dot{M} \omega R_{\text{m}}^2$  \\
         B & $\dot{M} \sqrt{2GMR_{\text{m}}}$ \\ 
         C & $\dot{M} \text{max}(v_\infty^2,~v^2_{\text{ff}}(R_\text{m})) / (2\omega)$ \\
         D & $\dot{M} v_\infty^2 / (2\omega)$ \\ \hline
    \end{tabular}
    \caption{Propeller models and corresponding spin-down torques $K$. Here, $v_\text{ff}(R_\text{m})=\sqrt{2GM/R_\text{m}}$. The authors of the models: A~--- \cite{1975SvAL....1..223S}, B~--- \cite{1973ApJ...179..585D}, C~--- \cite{1975AA....39..185I}, D~--- \cite{1981MNRAS.196..209D}.
    }
    \label{tab_prop}
\end{table}

 Eventually, the NS spins down and reaches the spin period corresponding to $R_\text{m}<R_\text{cb}$. We assume that it is the condition of the accretion onset, ignoring the possible subsonic propeller stage. In some rare cases with varying external parameters, i.e. if the external pressure decreases, the NS can switch back to the ejector or transient ejector stage. This means that the magnetosphere can expand to its maximum size $R_\text{m} = R_\text{l}$ and the NS can start generating the pulsar wind if $R_\text{Sh}^\text{env}>R_\text{l}$. If both conditions are satisfied, the spin-down proceeds with $K=K_\text{E}$.

With the onset of accretion, the way the NS interacts with the material changes, so the envelope can have a different profile. We assume that the standard BHL accretion starts, so the velocity of the material is close to the free-fall velocity $v_\text{ff}$ and the radius of the magnetosphere is $R_\text{A}$. The NS can both spin up and spin down, so the angular momentum consists of two terms
\begin{equation}
    K_\text{A} = K_\text{sd} - K_\text{su}.
\end{equation}
The spin-down torque is
\begin{equation}
    K_\text{sd} = k_\mathrm{t} \frac{\mu^2}{R_\mathrm{cb}^3},
\end{equation}
where coefficient $k_\text{t}=0.4$ \citep{2024Univ...10..205A}.
In the binary system, the material falling onto the NS has its angular momentum, which leads to the presence of the spin-up torque
\begin{equation}
    K_\text{su} = \dot{M} \eta \Omega R_\text{G}^2.
\end{equation}
If $K_\text{su}$ exceeds the torque required for disc formation around the NS $K_\text{disc}=\dot{M}\sqrt{GMR_\text{A}}$, then the spin evolution proceeds with $K_\text{su}=K_\text{disc}$. However, due to the low values of the orbital frequency $\Omega=\sqrt{G(M_\odot+M)/a^3}\approx 3\times 10^{-7}\text{~rad~s}^{-1}$, there is no disc in the systems considered.

As the external pressure decreases, the magnetosphere can expand beyond $R_\text{cb}$ and NS can switch back to the propeller stage. Since we are considering standard BHL accretion, the velocity of the matter near the magnetosphere is $v_\text{ff}$, and the radius of the magnetosphere is $R_\text{A}$. So the condition for the accretor-propeller transition is $R_\text{A}>R_\text{cb}$.

The stage switching can also be described in terms of the spin period $P$, so each of the considered transitions corresponds to a critical value of $P$. We dub it a transition period. For convenience, in the following, we will compare $P$ with the transition periods to determine the actual evolutionary stage, instead of considering the transition conditions written in terms of critical radii.

The spin evolution strongly depends on the value of the magnetic field, so it is important to consider how it changes with time. Generally, the field decays on several different timescales \citep{2021Univ....7..351I}. Throughout the evolution, the Ohmic dissipation is taking place. In addition, at early stages, the field dissipation can be enhanced due to the Hall cascade, which is especially important for the evolution of magnetars on the time scale $\tau_\text{Hall}\lesssim 10^5$--$10^6$~yrs.  
However, in our work, we do not focus on the variations within several million years, since the time interval for the evolution of the binary ($\sim 10$~Gyr) is much larger than $\tau_\text{Hall}$. 

We consider two models of the magnetic field evolution. In the first one, the magnetic field is assumed to be constant. 
In the second, the field decays exponentially during the whole evolution
\begin{equation}
  B = B_0 \exp(-t/\tau_{\text{Ohm}}).
\end{equation}
Here, $B_0$ is the initial value of the magnetic field. The time scale $\tau_\text{Ohm}$ is obtained in the following way. The typical initial magnetic field value of a radio pulsar is $\sim 10^{12}$~G. If we require it to decay to the lowest value observed in millisecond pulsars ($\sim 10^8$~G) within the Galactic age 13.6~Gyr, then $\tau_{\text{Ohm}} \approx 1.5 \times 10^{9}$~yr. This value is used in our calculations.

\subsection{The model of a binary system}
\label{sun_evo}

In this section, 
we first consider the main assumptions about the binary system used in our model. Then we define the evolution of the wind velocity $v_\text{w}$, the value of the speed of sound in the wind $c_\text{s}$, and the mass loss rate of the Sun-like star $\dot M_\text{w}$. These parameters are sufficient to determine all the necessary properties of the medium around the NS at a known radial distance from the companion. 

To specify the main assumptions in our model, we turn to observations. We focus on NSs in binaries similar to those observed by \textit{Gaia} and considered by \cite{2024A&A...686A.299S, 2024OJAp....7E..58E}. In general, the secondary components are Main sequence stars of masses $M_* = 0.7-1.4~M_\odot$, whose metallicity and age in some cases are not well known. According to \cite{2021LRSP...18....3V} this places them in a broad group of Sun-like (or solar-like) stars. We expect these stars to evolve independently of NSs since the components have large separations corresponding to long orbital periods $\gtrsim 1$~year. 
In this paper, we present calculations only for a companion of $1\, M_\odot$ with radius $R_\odot$, since properties and evolution of the stellar wind are much better known in the case of the Sun.
We focus on eccentric orbits with just a single value of the semi-major axis $a=1$~AU. This is a characteristic value for the observed systems and our calculations with circular orbits show that the dependence of the evolution of the NS on $a$ is weak \citep{2024arXiv240900714A}. 

We start our consideration of the main parameters of the wind with $v_\text{w}$. From the observational data obtained by Ulysses and ACE spacecrafts it is known that at $r\gtrsim1$~AU the solar wind consists of slow and fast components with mean velocities of $\sim400$ and $\sim760$~km~s$^{-1}$, respectively \citep{2015A&A...577A..27J}. We consider only the slow component of the wind because the fast wind has a much smaller impact on the evolution of the NS. This assumption is justified by the following facts. First, according to the standard BHL model, the accretion rate drops with increasing velocity, so there is much less matter captured from the fast wind. Second, the density of the fast wind is generally lower \citep{2005ApJ...623..511A}. Additionally, we not only exclude the fast component from the consideration, but we also ignore the fluctuations in the value of the wind velocity. This can be done as the main properties of the slow component remain relatively stable \citep{2021A&A...654A.111D}, despite the fact that the solar wind as a whole is anisotropic and varies during the solar cycle \citep{2003GeoRL..30.1517M}.

To define the value of $v_\text{w}$ and select the range of possible eccentricities $e$, we need to discuss the interval of radial distances between the two components that fits our model. We assume that the capture of the wind matter can be described by the standard BHL accretion with the characteristic distance $R_\text{G}$. According to this assumption, the NS is in the parallel flow of the incoming material with velocity $v$.  If the NS is too close to the second component (i.e., $r\lesssim R_\text{G}$) then this condition is violated. Thus, the minimal distance from the Sun-like star at which the NS captures matter in the BHL regime must satisfy the condition $r_\text{min}\gg R_\text{G}$. To estimate $r_\text{min}$, we first take $v_\text{w}=400$~km~s$^{-1}$. According to \cite{2005ApJ...623..511A, 2015A&A...577A..27J}, this is a typical value for the slow component of the wind at distances $r\gtrsim0.2$~AU from the star (we consider $r<0.2$~AU hereafter). We can estimate $R_\text{G}=2GM/v^2\sim 2GM/v_\text{w}^2\sim0.02$~AU. Here it is assumed that $v_\text{w}$ is considerably larger than $c_\text{s}$ and $v_\text{orb}$, which is true for $r>0.2$~AU. To satisfy the condition $r_\text{min}\gg R_\text{G}$, we set $r_\text{min}=10R_\text{G}$, so $r_\text{min}=0.2$~AU. In deriving this distance we assumed $v_\text{w}$ to be constant, but $v_\text{w}$ varies with the radial distance and this variation is significant when $r\lesssim0.2$~AU. However, this does not change the estimate of $r_\text{min}$, since for $r<r_\text{min}$ the wind speed $v_\text{w}$ decreases. This means that $R_\text{G}$ of the NS increases as it gets closer to the Sun-like star. So, the ratio $R_\text{G}/r$ can only increase when the wind velocity is larger than the orbital one. Therefore, we define that our model is only applicable to systems where the minimum distance between the NS and its companion is larger than $r_\text{min}=0.2$~AU. This estimate of $r_\text{min}$ also constrains the eccentricity $e$ because the periastron distance $a(1-e)$ should not be smaller than $r_\text{min}$, so $e_\text{max}=0.8$ for $a=1$~AU.
  
The next step is to consider the evolution of the wind. According to observations of hundreds of Sun-like stars \citep{2011ApJ...743...48W}, the stars generally spin down 
with time, which affects the activity of the star and, consequently, leads to a decrease in the mass loss rate. Here we need to define the evolution of the wind speed $v_\text{w}$ and number density $\rho$. Both of these parameters have been studied much less than the mass loss rate $\dot{M}_\text{w}$. 
Guided by observations, \cite{2015A&A...577A..28J} proposed two models of the wind evolution with the same $\dot{M}_\text{w}(t)$ which is constant for $t<300$~Myr and decreases at later time as
\begin{equation}
    \dot{M}_\text{w} = \dot{M}_{\text{w}0}\left(\frac{t}{4.6\text{~Gyr}}\right)^{-0.75},
\end{equation}
where $\dot{M}_{\text{w}0}=2\times10^{-14}\,M_\odot$~yr$^{-1}$ is the present-day mass loss rate of the Sun.

In one of the models, $v_\text{w}$ and $\rho$ both evolve with time. In the other model, only $\rho$ varies while $v_\text{w}$ remains almost constant. We adopt the latter model because in our consideration we are interested in the ram pressure $p$ and the accretion rate. Both are proportional to $\rho$ but have a stronger dependence on $v$: $p\propto v^2$ and $\dot{M} \propto v^{-3}$. Therefore, with the existing uncertainty in $\rho(t)$ and $v_\text{w}(t)$, choosing the model with constant $v_\text{w}$ would introduce a smaller error. Finally, we assume $v_\text{w}$ to be $400$~km~s$^{-1}$ for any age of the Sun-like star, while the number density of the wind is calculated from $\dot{M}_\text{w}$ via the continuity equation
\begin{equation}
\label{eq_cont}
    \rho(r) = \frac{\dot{M}_\text{w}}{4\pi r^2 v_\text{w}}.
\end{equation}

At considered radial distances, we expect $v_\text{w}>c_\text{s}$, so the exact value of the speed of sound in the medium is not so important. According to observations, a typical proton temperature at $1$~AU is $\sim10^5$~K (for example, observations considered by \cite{2019ApJ...887...83S, 2015A&A...577A..27J} to constrain the solar wind models show this value). The wind is typically isothermal \citep{1965SSRv....4..666P} or polytropic with $\Gamma$ near 1 \citep{2018MNRAS.481.5296V}. Therefore, we assume that the proton temperature is constant and equal to $10^5$~K. The corresponding sound speed is $c_\text{s}\approx\sqrt{k_\text{B}T/m_\text{p}}\approx30$~km~s$^{-1}$. 
I.e., much smaller than the wind velocity at $r\gtrsim 0.2$~AU. 



\subsection{Eccentric orbits}

As the position of an NS on the eccentric orbit changes, the parameters of the surrounding medium vary. This affects the evolution of the NS and may result in stage switching. We call transitions from ejector to propeller and from propeller to accretor as direct. Otherwise, the transition is called reverse.
 
\begin{figure}[t]
    \centering
    \includegraphics[width=\linewidth]{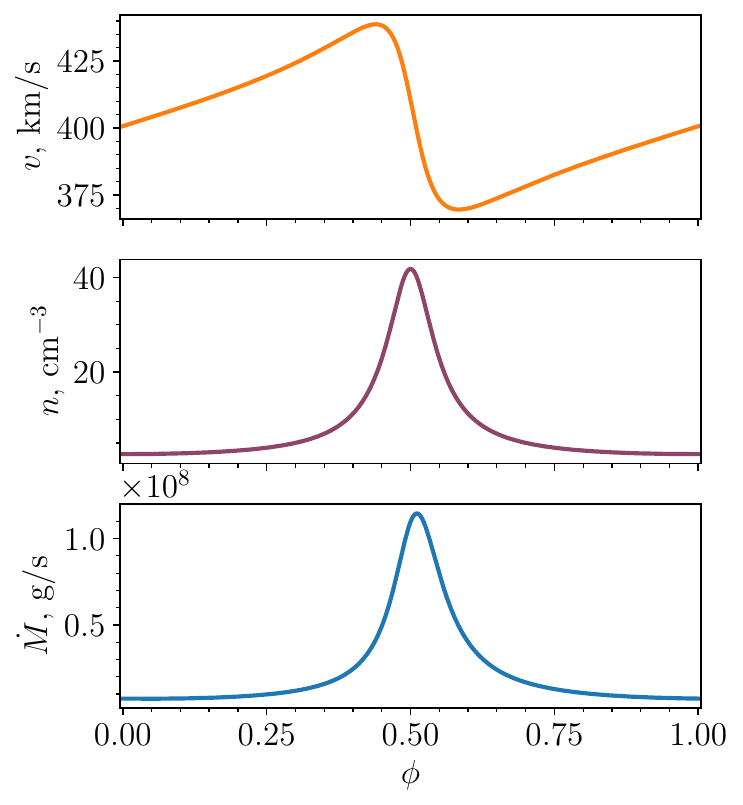}
    \caption{Variation of the velocity $v$ of the stellar wind relative to the NS, the number density of the external medium near the NS $n=\rho/m_\text{p}$ ($m_\text{p}$ is the proton mass), and the accretion rate $\dot{M}$ over the orbital phase $\phi$ (periastron corresponds to $\phi=0.5$). Here, the NS has the magnetic field $B=10^{12}$~G, the age of a Sun-like star is $4.6$~Gyr, the semi-major axis is $a=1$~AU, and the eccentricity is $e=0.6$.}
    \label{fig_orb}
\end{figure}

\begin{figure}[t]
    \centering
    \includegraphics[width=\linewidth]{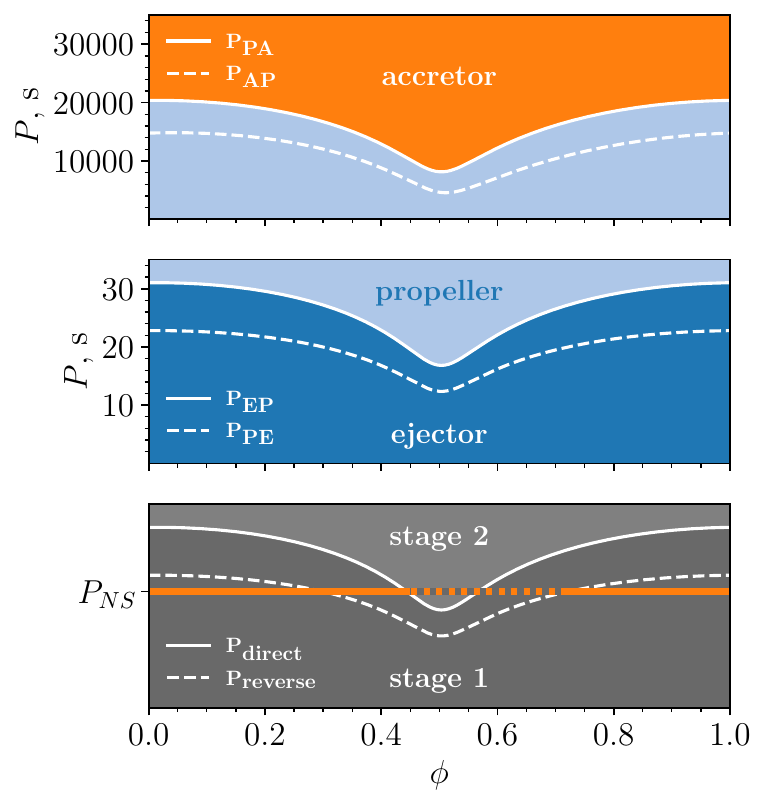}
    \caption{The changes in the transition spin periods of an NS with parameters similar to those in Figure~\ref{fig_orb}. The solid lines correspond to the direct transitions, i.e. the propeller-accretor transition (top panel) and the transition from the ejector to the propeller stage (middle panel). The dashed lines represent the reverse transition period. The bottom panel illustrates 
 the method for determining the evolutionary stages of an NS in an eccentric orbit (see the text). Here, an NS with a spin period $P_\text{NS}$ stays at the evolutionary stage 1 most of the time (solid orange line) and switches to stage 2 near periastron (dashed orange line).
    }
    \label{fig_per}
\end{figure}

First, we determine the variation of $\rho$, $v$, and $\dot{M}$ over an orbital period. An example of such changes for a typical system is shown in Fig.~\ref{fig_orb}. We then consider direct and reverse transitions between evolutionary stages for an NS in terms of critical spin periods. This is illustrated in Fig.~\ref{fig_per} and summarized later in this section.

First, we determine how the position of the NS in the orbit changes with time. This is the solution to Kepler's equation which cannot be found algebraically. Therefore, we use the Python package \verb|Gala| \citep{2017JOSS....2..388P} and numerically calculate the orbital velocity $\vec{v}_\text{orb}(\phi)$ and the radius vector $\vec{r}(\phi)$ of the NS. Here, $\phi$ is the orbital phase. It varies uniformly with time $\phi = t/T_\text{orb}$, where $T_\text{orb}$ is the orbital period. The origin of the coordinates corresponds to the position of the Sun-like star.

Then, we define how the key parameters~--- $\rho$ and $v$,~--- depend on $\vec{r}$ and $\vec{v}_\text{orb}$. 
The speed of the stellar wind relative to the NS is 
\begin{equation}
    \vec{v} = \vec{v}_\text{w} - \vec{v}_\text{orb}.
\end{equation}
Due to the orbital rotation, there would be two forces acting on the wind particles. However, since in our cases $v_\text{orb}$ is much smaller than $v_\text{w}$, we neglect it.


The density is calculated from eq.~(\ref{eq_cont}) with $r=|\vec{r}|$ and constant $c_\text{s}$ as shown in Sec.~\ref{sun_evo}. Finally, with known $\rho$, $v$, and $c_\text{s}$ we also determine the change in the accretion rate $\dot{M}$ during an orbital period.

Now, let us illustrate how an NS evolves in an eccentric orbit. It is important to note that the spin period of the NS does not change much during one orbital revolution, given that the orbital period is much shorter than the typical spin-down timescale. In other words, the spin period of the NS can be assumed to be constant during an orbital period, and it is the transition periods that vary. This variation can force the NS to change its evolutionary stage.

In our scenario, the NS is born at the ejector stage and spins down regardless of the external conditions, until the direct transition period $P_\text{EP}$ is reached. The first transition to the propeller stage is possible when the rising $P$ reaches the minimum value of $P_\text{EP}$ somewhere near the periastron, where the external pressure is higher. 
Now, for some time the NS switches between the propeller and ejector stages as it moves along the orbit.

As the spin period $P$ increases, so does the time spent at the propeller stage during one orbital revolution. Eventually, the period reaches the maximum period of the reverse transition to the ejector $P_\text{PE}$, making it impossible to return to the ejector stage even at the apoastron, when the external pressure is the lowest. Then, the NS stays in the propeller stage until it reaches the next transition period $P_\text{PA}$ near the periastron. The following picture with transitions between the propeller and accretor stages is similar. Finally, $P$ becomes long enough for the NS to accrete matter from the stellar wind throughout the whole orbital period. Due to the presence of the spin-up torque $K_\text{su}$ in addition to the spin-down moment $K_\text{sd}$, the NS can reach the equilibrium period $P_\text{eq}$, which is derived from the condition $K_\text{su} = K_\text{sd}$. We assume that the equilibrium period at the accretion stage is nearly constant on a large timescale, even though this formal equation yields different $P_\text{eq}$ at different points in the orbit because the timescale of spin-down or spin-up is typically much larger than an orbital period.

We need to account for stage switching due to the orbital motion in our modeling of the long-term spin behavior. 
We consider the evolution up to 10 billion years. This interval is divided into time steps $\Delta t=1$~Myr. 
Before the spin period $P$ reaches the minimum value of $P_\text{EP}$, calculations can be done rapidly as the NS always spins down according to eq.~(\ref{euler}) with $K_\text{E}$. As soon as the first transition to the propeller stage occurs, we need to proceed in a more complicated way.
Now, at each time step, we calculate if stage switching occurs and determine the values of $\phi$ where these transitions (direct and reverse) occur.

Let us illustrate this process with an example where we consider two stages (1 and 2), see the bottom panel of Fig.~\ref{fig_per}.  
The spin period $P$ is assumed to be constant during one orbital revolution. With the known $\rho(\phi)$ and $v(\phi)$, we determine if there are points in the orbit where the transitions occur. There can be only two points: for the direct and the reverse transitions. 

Note, that always $P_\text{direct}(\phi)>P_\text{reverse}(\phi)$.
If for all $\phi$ we have $P<P_\text{direct}(\phi)$, then the NS is at stage 1 along the whole orbit. In contrast, if for all orbital phases $P$ exceeds $P_\text{direct}(\phi)$ then it is at stage 2 for all $\phi$. Thus, in these cases, no stage switching occurs. 

Otherwise, there are direct and reverse transitions. We determine the phases, $\phi_\mathrm{d}$ and $\phi_\mathrm{r}$, at which they happen to move from $\phi=0$ (apoastron) along the orbit, as illustrated in the bottom panel of Fig.~\ref{fig_per}. The first (direct) transition corresponds to $P=P_\text{direct}(\phi_\mathrm{d})$ and the second (reverse) -- to $P=P_\text{reverse}(\phi_\mathrm{r})$.

\begin{figure}
    \centering
    \includegraphics[width=1\linewidth]{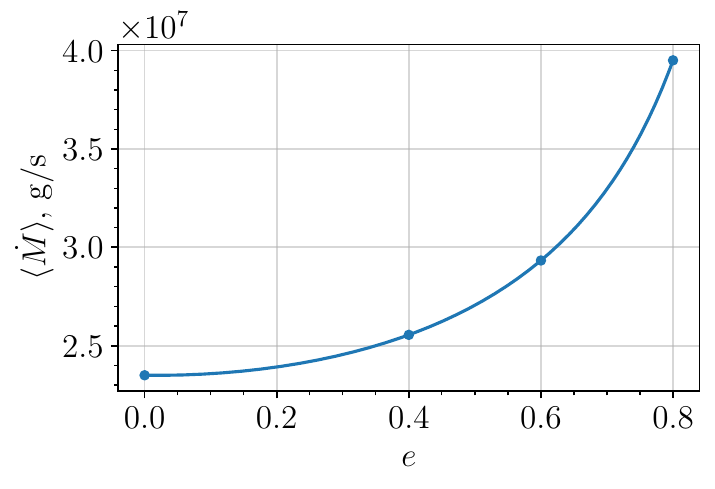}
    \caption{The orbit-averaged accretion rate of the NS in the orbit with an eccentricity $e$. The dots highlight the eccentricities we consider for the evolution of NSs. The value $\langle \dot{M}\rangle =\int_0^1 \dot{M}(\phi)\text{d}\phi$, the parameters of the Sun-like star correspond to an age of the Sun $4.6$~Gyr.}
    \label{fig_mdot}
\end{figure}

Once this is done, we calculate the change of the spin period at one integration step $\Delta t$ taking into account torque variation at different stages and the corresponding external conditions. The calculations done for a single orbit are scaled up to the whole time interval $\Delta t$, resulting in the increment of the spin period during one time step
\begin{equation}
    \Delta P = \Delta t\int_0^{1} K(\phi)\text{d}\phi,
\end{equation}
where the torque $K$ depends on the evolutionary stage and the behaviour of $v(\phi)$, $\rho(\phi)$, and $\dot{M}(\phi)$. In general, the eccentricity changes the mean value of $\dot{M}$, which is $\langle \dot{M}\rangle=\int_0^1 \dot{M}(\phi)\text{d}\phi$. This effect is shown in Fig.~\ref{fig_mdot}.
The age of the second component is also taken into account and considered equal to the age of the NS. 

\section{Results}
\label{sec_res}

Here we present our calculations of the evolution for an NS with a typical initial spin period $P_\text{0}=100$~ms \citep{2022MNRAS.514.4606I} in the orbit with the semi-major axis $a=1$~AU throughout the evolution of the second component on the Main sequence. We consider two initial values of the magnetic field of the NS $B_0$ -- the standard value $10^{12}$~G and the higher one $10^{13}$~G, -- and two models of its evolution. In one case, the field remains constant and in the other, it exponentially decays by $\approx4$~orders of magnitude in 10~Gyr. We apply models A and B of the spin-down at the propeller stage and vary the eccentricity of the orbits: $e=0$, $0.4$, $0.6$, and $0.8$. 

First, we examine how all these aspects separately affect the time of the accretion onset. For this reason, we have computed several scenarios, which are shown in Fig.~\ref{fig_12} for $B_0=10^{12}$~G and in Fig.~\ref{fig_13} for $B_0=10^{13}$~G. Then we calculate the time an NS spends at each evolutionary stage to estimate the probability of observing an accretor.

\subsection{Evolution of the NS with the standard magnetic field}

\begin{figure*}
	\centering 
	\includegraphics[width=1\textwidth]{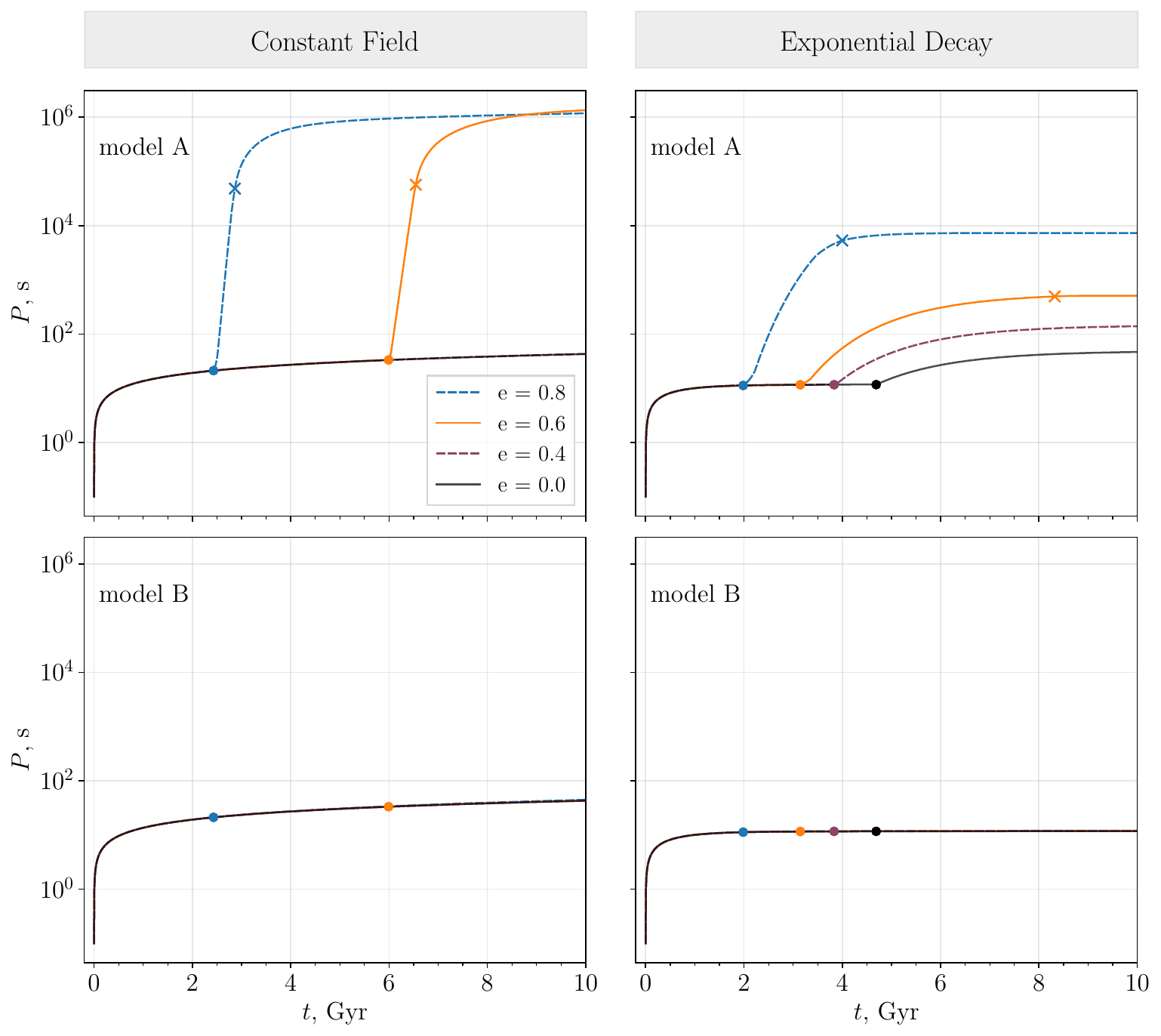}	
	\caption{The evolution of the spin period $P$ of an NS with initial magnetic field $B_0=10^{12}$~G in the orbit $a=1$~AU around a Sun-like star. On the left panels, the field is constant and equal to $B_0$, while on the right panels $B$ decays exponentially. The top and bottom panels show the evolution with two models of the spin-down at the propeller stage (models A and B, respectively). Line styles and colors are similar on all four panels. Each line corresponds to a different eccentricity of the orbit. Colored-filled circles show the first transition to the propeller stage. The crosses indicate the final transition to the accretor stage, which lasts for the entire orbital period.
    } 
	\label{fig_12}
\end{figure*}

Let us look at the spin period evolution of four NSs with a constant magnetic field $10^{12}$~G, within the propeller model that possesses the highest energy loss rates, model A, and with various eccentricities of the orbits (the upper left panel in Fig.~\ref{fig_12}). 

All NSs start as ejectors and lose rotational energy at the same rate, regardless of external conditions, until 2~Gyr. By this time, the NS with the highest $e=0.8$ has spun down enough to switch to the propeller stage 
near the periastron. Other NSs with lower $e$ have the same $P$, but they remain at the ejector stage. The reason is the following, they cannot get close enough to the companion star where the matter of the wind is the densest and the external pressure is as high as for the NS with $e=0.8$. In terms of transition periods, the minimum value of $P_\text{EP}$ (which is $P_\text{EP}$ at the periastron) is higher for them, so they have to spin down longer. As a result, the NS with $e=0.6$ switches to the propeller stage later, at 6.5~Gyr, while the other two NSs with $e=0$ and $e=0.4$ do not reach the propeller stage at all in 10 Gyr. 

After the transition, the NS continues to change stages twice every orbital period until it reaches the maximum value of $P_\text{PE}$ (which is $P_\text{PE}$ at the apoastron). Then the evolution continues with only the propeller stage until the NS reaches $P_\text{PA}$ near the periastron. The exact value of $P_\text{PA}$ also depends on the eccentricity, but since the spin-down within propeller model A is so effective, this difference does not significantly delay the transition to accretion. After another stage switching phase, the NS finally reaches the maximum value of $P_\text{AP}$ and can no longer switch back to the propeller stage. As accretion occurs throughout the whole orbital period, the spin period relaxes to the equilibrium spin period, which can be derived from the orbit-averaged equation $\langle K_\text{sd}\rangle =\langle K_\text{su}\rangle $. This period is longer for lower eccentricities. This can be explained as follows. The spin-down torque $K_\text{sd} = k_\mathrm{t} {\mu^2}/{R_\mathrm{cb}^3}\propto \omega^2 \propto P^{-2}$ does not contain any parameter varying with $\phi$, so $\langle K_\text{sd}\rangle =K_\text{sd}$. The spin-up torque $\langle K_\text{sd}\rangle = \langle \dot{M} \eta \Omega R_\text{G}^2\rangle \propto \langle \dot{M}\rangle $ ($R_\text{G}$ also changes with $\phi$, but the variation is much smaller, so we neglect it here), therefore $K_\text{su}$ is higher for more eccentric orbits. Since $P \propto K_\text{sd}^{-1/2}$ and $K_\text{sd}=K_\text{su}$, the equilibrium period is shorter for higher values of $e$. Thus, the NS in a more eccentric orbit reaches the next evolutionary stage earlier and has a slightly shorter equilibrium period as the accretion is established.

Now, let us demonstrate how the propeller spin-down rate affects the evolution. To do this, we compare the four binary systems considered previously with the same constant field $10^{12}$~G of the NS but with a less efficient propeller spin-down mechanism, i.e. propeller model B. The two left panels of Fig.~\ref{fig_12} can serve as an illustration here.

Once the first switch from the ejector to the propeller has occurred, the NS cannot immediately switch back to the ejector stage as the conditions for direct and reverse transitions are different. Thus, the NS will spend a significant portion of its orbital period at this stage, so its spin-down rate will be visibly affected by the choice of propeller model from the very first E-P transition. In the case of model A, the NSs begin to spin down dramatically immediately after reaching $P_\text{EP}$. In model B, the moments at which the NSs reach $P_\text{EP}$ are the same, but the spin periods do not change significantly afterward. This is because under the considered external conditions and with $B=10^{12}$~G, the spin-down in the propeller model B is comparable to that of the ejector stage. The same can be seen if we compare two propeller models for the NSs with exponentially decaying magnetic field and the same initial value $B_0=10^{12}$~G (right panels in Fig.~\ref{fig_12}). While in model A the spin period changes visibly after the start of the propeller stage, in model B it remains almost constant. This excludes the transition to the accretion stage within 10~Gyr. 

In this paper, we do not present calculations for models C and D, as they do not lead to the accretion stage even for high values of $e$ and $B_0$. Thus, the choice of the spin-down at the propeller stage can easily delay the onset of accretion or even make it impossible.

The next important aspect of our modeling is the evolution of the magnetic field. This can be illustrated by comparing the two left panels with the two right panels in Fig.~\ref{fig_12}. 
Before the accretion onset, the magnetic field decay influences the evolution of the NS in two ways. On the one hand, lower values of $B$ result in less efficient spin-down at both the ejector and the propeller stages, so that the dependence $P(t)$ is visibly less steep as the field decays with time. Thus, the spin evolution is slowed down in this case. On the other hand, the decrease in $B$ causes the magnetosphere radius to shrink, which makes the transition to the next stage easier by shortening the corresponding transition period. The calculations show that the latter effect is more important for reaching the propeller stage. 
For example, the ejector stage of NSs with the constant field $B=10^{12}$~G in the orbits with $e=0.4$ and $e=0$ lasts longer than 10~Gyr. 
In contrast, the same NSs with a decaying field switch to the propeller stage in $\sim $~a few~Gyr. The difference in transition periods also affects the P-A transition, but here the effect of the slower spin-down rate seems to be more significant. Thus, even though the NS has to reach a lower value of $P_\text{AP}$, the transition to accretion, which would last the entire orbital period, is delayed.

Now we consider how the choice of the field evolution model affects the accretor stage. Above, we have shown that the spin period $P$ is equal to the equilibrium period $P_\text{eq}$, so it is longer for lower $e$, but this is only true if the field is constant. Our calculations show that it is the opposite for a decaying field. 
In fact, $P_\text{eq}$ decreases with time along with decreasing $B$. However, the spin period of the NSs seems to be almost constant at the accretor stage. This is so because to maintain $P=P_\text{eq}$ the NS should have spun up due to $K_\text{su}$ but it has no time to do so. In other words, the typical spin-up timescale $\tau_\text{su}$ here is longer than the timescale of changes in $P_\text{eq}$, which is governed by $B(t)$ and therefore similar to $\tau_\text{Ohm}$. So, the NS keeps the spin period gained at the previous propeller stage. As the NS with the highest $e$ reaches the propeller stage earlier, it has a larger $B$ at this moment and therefore a longer $P_\text{AP}$. Thus, the decaying magnetic field helps to reach the propeller stage, but the accretor stage is delayed. After the stage of accretion is reached, the way the field decays has little impact on the subsequent spin evolution.

\subsection{The evolution with a higher magnetic field}

\begin{figure*}
	\centering 
	\includegraphics[width=1\textwidth]{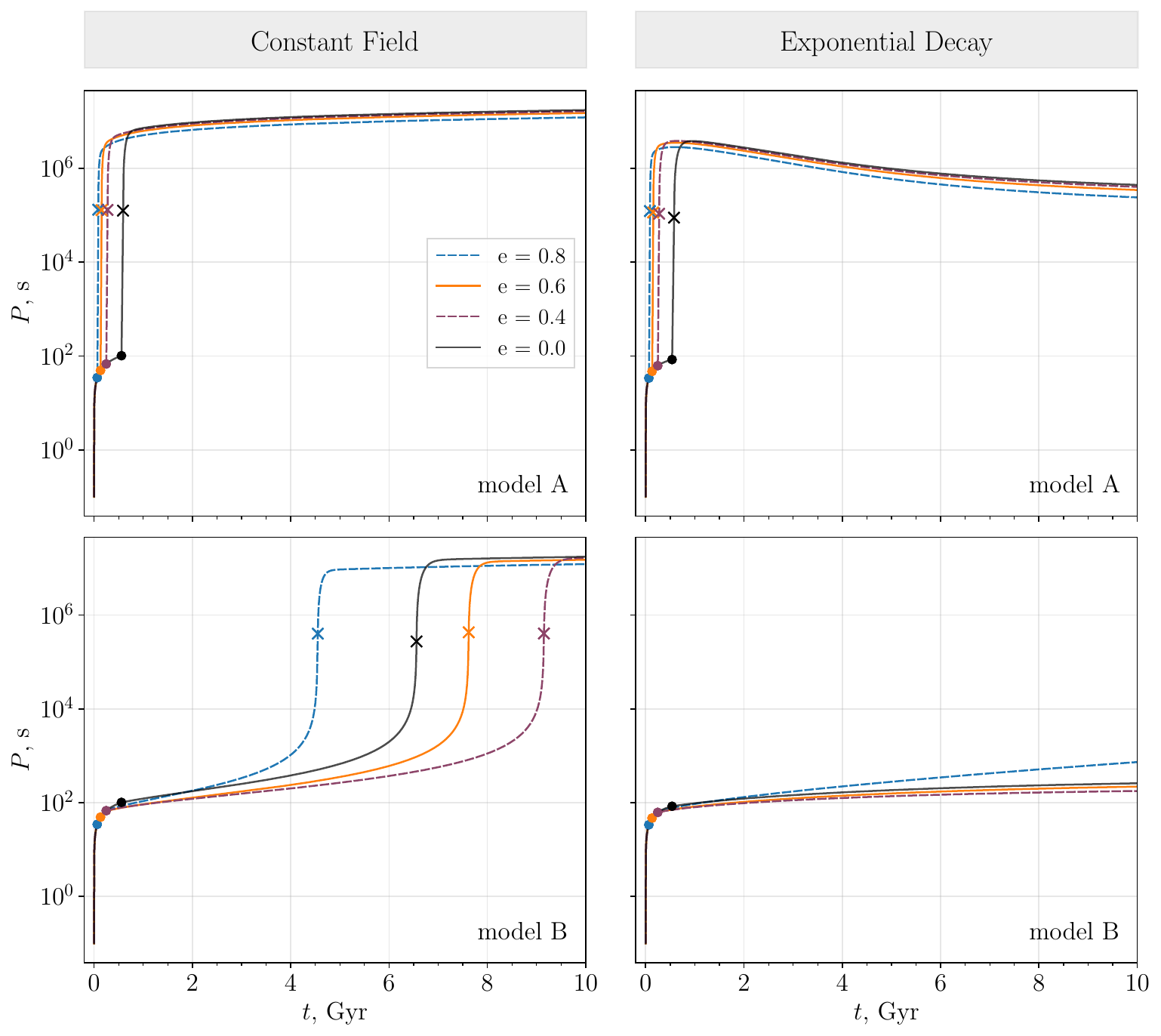}	
	\caption{Spin evolution of an NS with a high initial magnetic field value, $B_0=10^{13}$~G, orbiting around a Sun-like star, the semi-major axis is $a=1$~AU. The symbols, line styles, and colors are similar to those in Figure~\ref{fig_12}.
    }
	\label{fig_13}%
\end{figure*}

The spin evolution of NSs before accretion strongly depends on the magnetic field value. That is why it is important to consider different values of $B_0$. In the following analysis in this subsection, we accept $B_0=10^{13}$~G and refer to Fig.~\ref{fig_13} comparing it with Fig.~\ref{fig_12}.

Most importantly, in all models of propeller spin-down and magnetic field decay, the NSs with the higher value of $B_0$ evolve faster. This results in shorter stages of the ejector and propeller and also leads to an earlier onset of accretion. For example, the ejector stage for some of the NSs with $B_0=10^{12}$~G lasts for more than $10$~Gyr, while with $B_0=10^{13}$~G the NSs switch to the propeller stage within the first billion years. In both models (A and B) of the propeller spin-down, the subsequent transition to the accretor stage is earlier when $B_0$ is higher. In model A, the transition to accretion is almost instantaneous after the propeller stage is reached. The effect within model B can best be seen by comparing the two magnetic field values of the NSs with the constant field. With $B_0=10^{12}$~G the spin period is almost constant after reaching the propeller stage, while for $10^{13}$~G it increases by several orders of magnitude, allowing all NSs to begin accreting within $10$~Gyr.

Overall, the conclusions made for the NSs with a standard magnetic field are also applicable to the NSs with $B_0=10^{13}$~G; however, several adjustments must be made. We noted earlier that a higher $e$ leads to an earlier transition to the propeller stage, and this is also true for $B_0=10^{13}$~G. However, the subsequent onset of accretion does not always undergo the same pattern if the propeller spin-down is inefficient. Within propeller model B, the NSs with the constant $B=10^{13}$~G enter the propeller stage for the first time earlier for the highest $e$ and later for the lowest. However, the NS with $e=0$ starts to accrete later than the one with $e=0.8$, but earlier than $e=0.6$.
This happens because the spin-down rate of the ejector with $B=10^{13}$~ G is higher than that at the propeller stage in model B. So, on the one hand, a higher $e$ leads to a higher $\dot{M}$, which makes the NS switch the stage to the next earlier. On the other hand, the lowest eccentricity allows the NS to spin down more in the same time interval. The interplay of these two effects causes the NSs to start accreting out of order.

\subsection{The time spent at different evolutionary stages}

The probability of observing an accretor depends on the fraction of time the NS can accrete. For this reason, we calculate the time spent at each evolutionary stage for NS within two propeller spin down models, A and B, with two values of the constant field, $10^{12}$~G and $10^{13}$~G, and four values of eccentricity, as shown in Table~\ref{tab_res}. 

The time spent at the accretor stage $t_\text{A}$ ranges from 0 to $\sim100$\% regarding the chosen model of the spin evolution. In general, NSs spend more time at the accretor stage $t_\text{A}$ if their magnetic field and the eccentricity of their orbits are higher and if the spin-down rate at the propeller stage is more efficient. For example, the eccentricity of the orbit determines $t_\text{A}$ for $10^{12}$~G. Thus, the NS with $B_0=10^{12}$~G can spend up to 70\% of its time as an accretor if $e$ is high enough, while it will not start accreting if the orbit is circular. If the magnetic field is high ($B_0=10^{13}$~G) and the propeller stage does not delay the onset of accretion much, we have almost 100\% chance of observing an accretor. In the case of the standard magnetic field and the inefficient propeller, $t_\text{A}$ is zero regardless of $e$. However, a significant chance $\approx10-50$\% (depending on $e$) of observing an NS at the accretor stage remains when the magnetic field is high. Thus, different components of the spin evolution of NSs lead to very different values of $t_\text{A}$. 


\begin{table}
\begin{tabular}{c|cccc|cccc} 
 
B & \multicolumn{4}{c|}{$10^{12}$~G} & \multicolumn{4}{c}{$10^{13}$~G} \\ 
e & 0 & 0.4 & 0.6 & 0.8 & 0 & 0.4 & 0.6 & 0.8 \\ 
\hline
$E_\text{A}$ & 1.00 & 1.00 & 0.61 & 0.26 & 0.06 & 0.03 & 0.02 & 0.01 \\
$P_\text{A}$ & 0.00 & 0.00 & 0.06 & 0.04 & 0.01 & 0.01 & 0.01 & 0.01 \\
$A_\text{A}$ & 0.00 & 0.00 & 0.34 & 0.70 & 0.93 & 0.97 & 0.98 & 0.98 \\\hline
$E_\text{B}$ & 1.00 & 1.00 & 0.92 & 0.89 & 0.06 & 0.03 & 0.03 & 0.02 \\
$P_\text{B}$ & 0.00 & 0.00 & 0.08 & 0.11 & 0.61 & 0.88 & 0.74 & 0.44 \\
$A_\text{B}$ & 0.00 & 0.00 & 0.00 & 0.00 & 0.34 & 0.09 & 0.24 & 0.54 \\
\end{tabular}
\caption{The fraction of time that an NS with a constant magnetic field spends in each stage~--- ejector (E), propeller (P), accretor (A)~--- during its 10~Gyr evolution on the orbit $a=1$~AU around a Sun-like star. Subscripts A and B correspond to the propeller models.
}
\label{tab_res}
\end{table}

\section{Discussion}
\label{disc}


In this section, we discuss our assumptions and some uncertainties of the model.

\subsection{Diversity of the stellar and binary parameters}

 We presented calculations for a very limited set of parameters of binary systems.  
Already in the known sample of `dormant' NSs in wide binaries with low-mass companions we see a variety of stellar masses, ages, orbital separations, etc. 
In addition, the parameters of the NSs can also vary. 
 Also, in our calculations, we do not model short-period systems. However, a few of them are already known thanks to spectroscopic observations. For example, two systems discussed by \cite{2024A&A...686A.299S} have orbital periods $<1$~day. 
In this subsection, we justify our choice.

Calculations for broad realistic distributions of various parameters are much welcomed in the framework of the population synthesis approach. Detailed accounts for the preceding evolution of binary systems must allow for the determination of the parameters, particularly for survived binaries. This depends, in particular, on the mass loss from the system and the kick velocity imparted to the NS. We plan to perform such modeling in the future. 

 Population synthesis studies of the magneto-rotational evolution of NSs in wide low-mass binaries have not been performed, yet. In this note, we make the first step by presenting our method of calculating the spin evolution of an NS for a restricted sample of parameters.  
 
Modeling the evolution of NSs in very tight systems is more challenging in many respects. A different approach is required to calculate the parameters of the external medium in the vicinity of a compact object. Stellar wind properties at small distances are much more variable and less known. These uncertainties might influence the final results significantly. In addition, stars in systems with small orbital separations have passed through an episode of very intense interaction, e.g. a common envelope stage. As in the present study, we do not model the binary evolution, we postpone studies of tight systems for the future.

Finally, we use just one value of the initial spin period, two values of the initial magnetic field of the NSs, and two (most simple) models of the field evolution. This allows us to illustrate the evolution of the most typical NSs. Despite we do not model the evolution of magnetars explicitly, their behavior is expected to be very similar to the studied cases. The large magnetic field of magnetars rapidly decays down to the standard values $\lesssim 10^{13}$~G. The characteristic time scale of the evolution of a strong field is much smaller than the duration of the ejector stage, see e.g., \cite{2021Univ....7..351I} for a review. Thus, rapid initial spin down can be mimicked by an increase in the initial spin period up to $\sim 10$~s. However, this does not influence the long-term evolution in the sense of transitions to the subsequent stages (propeller and accretor) in the case of low-density surroundings, see \cite{2024PASA...41...14A}.

\subsection{Model of the magneto-rotational evolution}

 Our main goal is to determine the moment when an NS can start to accrete. 
That is why we are mainly interested in the magneto-rotational evolution prior to the stage of accretion. There are many uncertainties related to the spin-down of NSs at the ejector and propeller stages. 

 It is important to underline that the ejector stage is much longer than the interval of time when the NS can be observed as a radio pulsar. The standard assumption for the spin-down rate at this stage is shown in eq.~(\ref{euler}) with $K=K_\text{E}$. There are several thousand known radio pulsars (see the ATNF catalog, \cite{2005AJ....129.1993M}), but even for these sources the validity of eq.~(\ref{K_E}) is somehow doubtful. For the rest of the ejector stage, we just do not have a single object to constrain the rate of the spin evolution. One has to rely on either theory or indirect estimates from the evolution of accreting NSs in binary systems.   

Note that the duration of the ejector stage in wide low-mass binaries is estimated to be $\sim$~a few Gyr for the standard magnetic field. Thus, there is a significant probability of catching an NS at such a stage. If it is possible to determine the spin period of such an object (with a known age that can be determined from the age of the optical companion), then it would be of great importance, as it would be a unique possibility to probe the spin-down rate at the ejector stage. 

In this paper, we applied several models for the propeller spin-down. 
The time of accretion onset for low $\dot M$ strongly depends on assumptions about the propeller stage. This is similar to the case of the evolution of isolated NSs; see, e.g., \cite{2024PASA...41...14A}. Due to the low density of the external medium, the efficiency of the spin-down mechanism at the propeller stage can be very low. For some of the proposed models (e.g. \citet{1975AA....39..185I, 1981MNRAS.196..209D}), an NS in a wide low-mass binary (or in the case of an isolated object) never reaches the accretor stage. In this case, the spin period is nearly 'frozen' at values $\sim$~a few tens of seconds. It would be rather difficult to detect any emission from an NS at such a stage because the expected luminosity in all ranges is very low. 

\subsection{Influence of variations in stellar parameters}

{ 

We have considered the NS evolution when the second component is almost identical to the Sun. However, observational data \citep{2024OJAp....7E..58E, 2024A&A...686A.299S} suggest that these stars differ in their parameters. In this subsection, we discuss how the differences in mass, radius, and metallicity of a Sun-like star can affect the evolution of the NS.

First, we illustrate how the mass-loss rates vary depending on radii $R_*$ and masses $M_*$ of the companions. According to the mass-radius diagram presented by \cite{2021A&ARv..29....4S}, for stars with $M_* = 0.7-1.4\,M_\odot$ the mass-radius dependence is close to $R_*\propto M_*$. Based on observations, \cite{2015A&A...577A..28J} have estimated how the mass-loss rate scales with mass, radius, and the rotational frequency $\Omega_*$ of the Sun-like star

\begin{equation}
    \label{johnstone}
    \dot{M}_\text{w} \propto R_*^2M_*^{-3.36}\Omega_*^{1.33}.
\end{equation}
The rotational frequency does not change significantly with mass.
\cite{2020ApJ...889..108A} have modelled the evolution of stars with $M_*=0.7,~1.0$, and $1.4\,M_\odot$ and have shown that the values of $\Omega_*$ are similar on the long timescale of $\sim 10^9$~yr. So we can ignore the difference in $\Omega_*$ for stars of the same age and different masses and rewrite eq.~(\ref{johnstone}) as $\dot{M}_\text{w} \propto M_*^{-1.36}$. Considering that the asymptotic speed of the slow wind $v_\text{w}=400\,$km~s$^{-1}$ does not change with mass, the accretion rate onto the NS has the same dependence on the stellar mass $\dot{M}\propto M_*^{-1.36}$. If the mass of the star is in the range $0.7\,M_\odot<M_*<1.4\,M_\odot$, we conclude that $\dot{M}$ can only be $\approx1.6$ times higher or lower than the one considered in Sec.~\ref{sec_res}.

Metallicity is another important factor that can influence $\dot{M}_\text{w}$. The data obtained by \cite{2024OJAp....7E..58E} suggest that there are several systems with metallicities a few tens times lower than the solar one. In a theoretical study, \cite{2018PASJ...70...34S} shows that 
the mass-loss rate for such a range of metallicity would be $\approx2.3$ times higher than for the Sun. 

In Fig.~\ref{fig_dis}, we illustrate how these changes in $\dot{M}$ affect the evolution of the NS with constant magnetic field $B=10^{12}$~G in the orbit with $a=1$~AU and $e=0.8$. For reasonable values of masses, radii, and metallicities the differences in the mass-loss rate can hardly exceed half an order of magnitude. For this reason, we present the evolution of the NS for three values of $\dot{M}$: the one used in Sec.~\ref{sec_res}, three times higher, and three times lower. Generally, these variations in $\dot{M}$ do not change the results qualitatively. However, quantitatively lower metallicities and masses $M_*$ result in a faster evolution of the NS.

\begin{figure}[t]
    \centering
    \includegraphics[width=\linewidth]{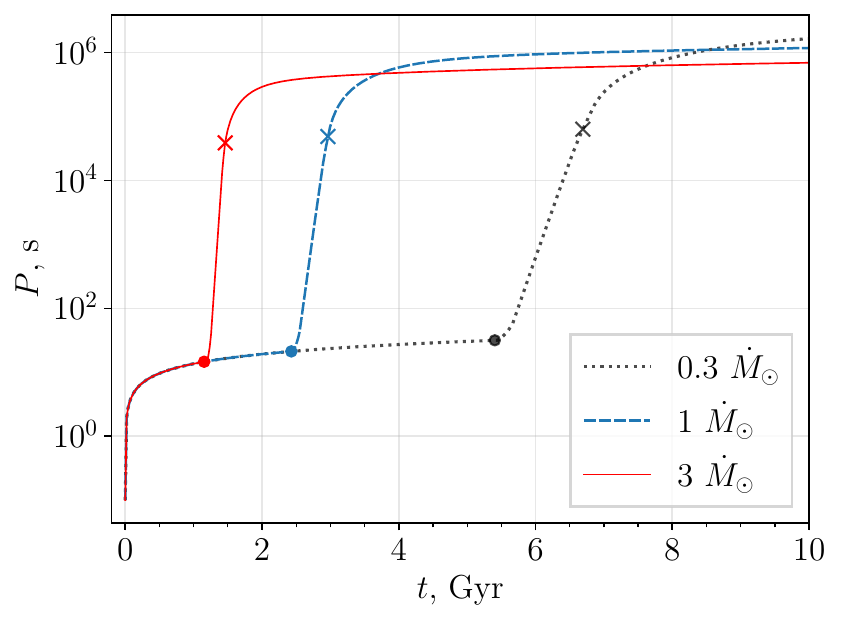}
    \caption{The spin evolution of the NS with the constant magnetic field $B=10^{12}$~G, $e=0.8$, $a=1$~AU in the case of the propeller model A. The lines are plotted for three values of the accretion rate which is given in units of the rate corresponding to the Sun as the donor, $\dot M_\odot$. The symbols are similar to those in Figs.~\ref{fig_12}, \ref{fig_13}. The curve for $1\,\dot{M}_\odot$ is exactly the same as the curve for $e=0.8$ in the upper left panel of Fig.~\ref{fig_12}.}
    \label{fig_dis}
\end{figure}

Note, that by varying the semi-major axis $a$ or/and eccentricity we can obtain a more significant influence on the evolution of an NS in a binary. Thus, we conclude that for realistic ranges of masses and metallicities, our results remain a robust illustration of the magneto-rotational evolution of an NS.   

 The Main sequence lifetime of a star depends on its metallicity, too. For lower metallicity, the Main sequence for sun-like and lower-mass stars is shifted towards the left and up in the Hertzsprung-Russel diagram. Thus, with a higher luminosity and temperature a star has a shorter lifetime in the Main sequence. The effect can be illustrated using stellar evolution calculations. E.g., in \verb|PARSEC| v1.2 tracks for a solar mass star the core hydrogen burning time is $\approx 9.08$ Gyr for $Z=0.020$, $\approx 8.65$ Gyr for $Z=0.017$, and just $\approx 4.95$ Gyr for $Z=0.001$ \citep{2012MNRAS.427..127B}. Thus, despite the fact that the NS evolution around a low-metallicity star proceeds faster due to a more intensive mass loss, the available time for this evolution is shorter. 
After the donor becomes a red giant, the NS can rapidly switch to the stage of accretion and, probably, appear as a symbiotic X-ray binary \citep{2019MNRAS.485..851Y}. 
}

\subsection{Observability of neutron stars}

 An NS in a wide low-mass binary can be either an ejector, a propeller, or an accretor. In all the cases, they might be dim sources. 

 At the ejector stage, one can expect some interaction of the relativistic wind from the NS with the wind from the Sun-like companion. The power of both winds is only $\sim 10^{27}$--$10^{28}$~erg~s$^{-1}$. Any emission related to the ejector wind could be modulated at the spin period of the NS $\sim 10$~s. This might be helpful. Still, the expected flux is very low. 

 For the propeller stage, a rough estimate of luminosity can be obtained from the magnetospheric accretion:
 \begin{equation}
 L_\text{prop}= \frac{GM}{R_\text{m}}\dot M\approx 10^{23} \dot M_8   \left(\frac{M}{M_\odot}\right)\left( \frac{R_\text{m}}{10^{11}\, \mathrm{cm}}\right)^{-1}\text{~erg~s}^{-1}.
 \end{equation}
Some energy can be also deposited due to the spin-down of the NS. Still, propellers in systems with small $\dot M$ are expected to be elusive sources. 

 The situation with an accreting NS is less clear, but also not very optimistic. The upper limit for persistent luminosity can be derived as 
 \begin{equation}
 L=\frac{GM}{R_\text{NS}}\dot M\approx 10^{28} \dot M_8 \left(\frac{M}{M_\odot}\right)\left( \frac{R_\text{NS}}{10^{6}\, \mathrm{cm}}\right)^{-1} \text{~erg~s}^{-1} .
 \end{equation}
 However, as in the case of isolated NSs, a realistic luminosity can be much lower than this value, or the accretion rate on the surface can be transient, see \cite{2015MNRAS.447.2817P}. 

 The accretion regime at low rates is poorly constrained by observations. Theoretical studies are not conclusive. Thus, if propeller spin-down is efficient enough, NSs in wide low-mass binaries can provide a unique opportunity to probe small $\dot M$ accretion. 



\section{Summary and conclusions}
\label{conc}

In this paper, we have studied the long-term evolution of NSs in wide eccentric binary systems with Sun-like companions. The parameters of the binaries correspond to those determined for several systems discovered by {\it Gaia}, see \cite{2024OJAp....7E..58E, 2024A&A...686A.299S}. We considered orbits with different eccentricities and analyzed the behavior of NSs with a standard initial magnetic field $10^{12}$ and a higher field $10^{13}$~G, applying several models of spin-down at the propeller stage and magnetic field decay. 

We have demonstrated that the eccentricity of the orbit plays an important role in the evolution of the NS: the ejector stage ends much earlier for NSs in orbits of higher eccentricity. Contrary to the case of NSs with $B_0=10^{12}$~G in circular orbits considered by \cite{2024arXiv240900714A}, the NSs with the same field but in eccentric orbits can switch to the propeller stage within the lifetime of their companions. Thus, they can start accreting within 10~Gyr  (i.e., the lifetime of the second component) if the propeller spin-down mechanism is efficient enough, e.g., model A by \citet{1975SvAL....1..223S}.

Overall, NSs with more eccentric orbits and higher magnetic field values make the transition to the propeller stage earlier. The subsequent transition to accretion depends strongly on the model of the propeller spin-down, while the evolution of the magnetic field is much less important. Thus, in the two efficient models, A and B, proposed by \citet{1975SvAL....1..223S} and \citet{1973ApJ...179..585D}, NSs with $B_0=10^{12}$~G (for $e\gtrsim0.6$) and $10^{13}$~G spend a few Gyr accreting stellar wind and potentially, can be observed e.g., as X-ray sources. Otherwise, the NS spends most of its life in the propeller stage with a spin period of $\sim$~a few tens of seconds.


\section*{Acknowledgements}
We thank Prof. N.I. Shakura for discussion. MA acknowledges support from the Basis Foundation (grant 24-2-1-39-1). We also thank the referee for the useful comments.

\bibliographystyle{elsarticle-harv} 
\bibliography{ecc_orb}

\end{document}